\documentclass[preprint2]{emulateapj}
\usepackage{multirow}

\newcommand{\kms}{\ensuremath{\rm{km\,s^{-1}}}}

\newcommand{\dwarf}{d0944$+$71}

\shorttitle{}
\shortauthors{Toloba et al.}

\begin{document}

\title{Spectroscopic confirmation of the dwarf spheroidal galaxy
  \dwarf\ as a member of the M81 group of galaxies}


\author{Elisa~Toloba\altaffilmark{1,2}}\email{toloba@ucolick.org}
\author{David~Sand\altaffilmark{1}}
\author{Puragra~Guhathakurta\altaffilmark{2}}
\author{Kristin~Chiboucas\altaffilmark{3}}
\author{Denija~Crnojevi\'c\altaffilmark{1}}
\author{Joshua~D.~Simon\altaffilmark{4}}

\affil{$^1$Texas Tech University, Physics Department, Box 41051, Lubbock, TX 79409-1051, USA}
\affil{$^2$UCO/Lick Observatory, University of California, Santa Cruz, 1156 High Street, Santa Cruz, CA 95064, USA}
\affil{$^3$Gemini Observatory, 670 North A’ohoku Pl, Hilo, HI 96720, USA}
\affil{$^4$Carnegie Observatories, 813 Santa Barbara Street, Pasadena, CA 91101, USA}

\begin{abstract}

We use Keck/DEIMOS spectroscopy to measure the first 
velocity and metallicity of a dwarf spheroidal (dSph) galaxy beyond the Local
Group using resolved stars. Our target, \dwarf, is a faint dSph found in the halo
of the massive spiral galaxy M81 by Chiboucas et al. We coadd the
spectra of 27 individual stars and measure a
heliocentric radial velocity of $-38\pm10$~\kms. This velocity is
consistent with \dwarf\ being gravitationally bound to M81. We coadd
the spectra of the 23 stars that are consistent with being red giant
branch stars and measure an overall metallicity of
${\rm [Fe/H]}=-1.3 \pm 0.3$ based on the calcium triplet lines. This metallicity is consistent with
\dwarf\ following the
metallicity$-$luminosity relation for Local Group dSphs. We investigate several potential sources of observational bias but find that our sample of targeted stars is representative of the metallicity distribution function of \dwarf\ and any stellar contamination due to seeing effects is negligible. The low ellipticity of the galaxy and its position in the
metallicity$-$luminosity relation suggest that \dwarf\ has not been
affected by strong tidal stripping.

\end{abstract}

\section{Introduction}

Dwarf galaxies play a critical role in our understanding of galaxy
formation in the context of the $\Lambda$ Cold Dark Matter
($\Lambda$CDM) paradigm of structure formation.  For instance,
quantitative verification of the $\Lambda$CDM model has struggled in
the dwarf-galaxy regime (e.g. the ``missing satellites problem",
\citealt{Klypin99,Moore99}; the ``too big to fail problem", \citealt{BoylanKolchin11,BoylanKolchin12}; and apparent planes of satellites, \citealt{Ibata13,Pawlowski15}), although the most recent numerical simulations make significant progress on several of these issues by including a wide range of baryonic physics \citep[e.g.][]{Brooks14,Wetzel16}.  Additionally, the low mass and large numbers of faint dwarf galaxies make them good targets to learn about environmental processes (e.g. tidal and ram pressure stripping), which also makes them vital contributors to the build-up of massive halos \citep[e.g.][]{Johnston08}.  Finally, the  star formation histories of Local Volume dwarf galaxies can push the high redshift ultraviolet luminosity function to fainter limits than can direct high redshift constraints  \citep[e.g.][]{Weisz14}.

The challenges to the $\Lambda$CDM model on small scales in particular have been largely based on studies in the Local Group, even though a large dispersion in the numbers and properties of dwarf satellites and other halo substructures are expected \citep[e.g.][]{Johnston08,Busha10}.  Recent work has thus sought to push the study of faint dwarf galaxies to other Local Volume galaxies \citep[e.g.][among others]{Chiboucas09,Chiboucas13,MD15,Merritt14,Sand15,Carlin16}, and our own team is conducting the Panoramic Imaging Survey of Centaurus
and Sculptor (PISCeS) program in the halos of
NGC~5128 and NGC~253 \citep{Sand14,Crnojevic14,Crnojevic16,etj16a}.  

These studies of faint dwarf galaxy satellites outside the Local Group have largely been photometric in nature, unless there is an accompanying HI neutral gas detection \citep[e.g.][]{Roychowdhury12,Sand15} or considerable effort is expended to obtain an integrated light spectrum \citep[e.g.][]{vandokkum15,vandokkum16}.  Spectroscopic information, such as a line of sight velocity and mean metallicity measurement, would add great value to these programs.  For instance, dwarf galaxy velocities will be invaluable for investigating the orbital history of satellite systems, the overall halo mass of the primary galaxy and claims of satellite planes.  Metallicity measurements would allow for studies of the luminosity -- metallicity relation in new environments \citep{Kirby11}.

We have devised a new method to obtain critical spectroscopic information via resolved stellar populations out to distances of $\sim 4$~Mpc, utilizing powerful ground-based multi-object spectrographs \citep[as recently described in][]{etj16c}.  The method involves co-adding spectra from a single dwarf galaxy or stellar stream to obtain a final spectrum to measure a radial velocity or mean metallicity.  Individual slits are placed on carefully chosen stars that are spatially associated with the targeted stellar structure --  tip of the red giant branch (TRGB) stars, asymptotic giant branch (AGB) stars and apparent stellar blends -- so as to maximize the final signal to noise of the co-added spectrum.  

This is the first of a series of papers in which
we will analyze the dynamical and metallicity properties of dwarf
galaxies and streams that reside in the halos of massive galaxies 
beyond the Local Group.  The spectroscopic subject of this paper is
\dwarf, which was discovered by \citet{Chiboucas09} during a CFHT
search for faint dwarf galaxies around M81, and was subsequently
confirmed to be at the distance of M81 \citep[$D=3.63$~Mpc;][]{Karachentsev02} with follow-up {\it Hubble Space Telescope} observations \citep{Chiboucas13}.  We present several previously known physical properties of \dwarf~ in Table~\ref{data}; it has an absolute magnitude of $M_{I}$=$-$13.2, is gas poor \citep[$M_{HI}$$<$3.1$\times$10$^5$ $M_{\odot}$;][]{Roychowdhury12}, and lies at a projected distance of 335 kpc from M81 itself.

\section{Data}

\subsection{Observations and Data Reduction}\label{observations}

\begin{figure*}
\centering
\includegraphics[angle=0,width=8.5cm]{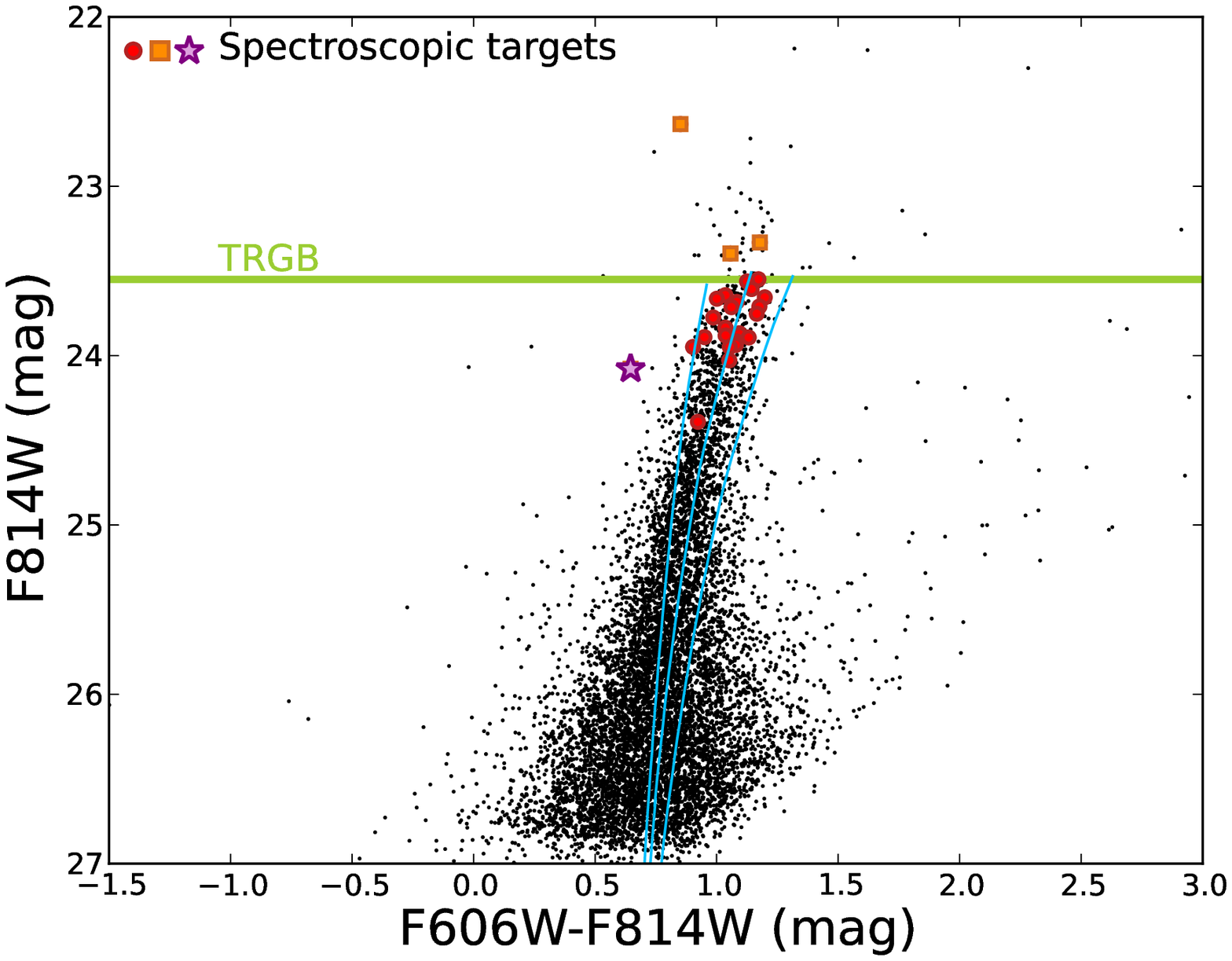}
\includegraphics[angle=0,width=8.5cm]{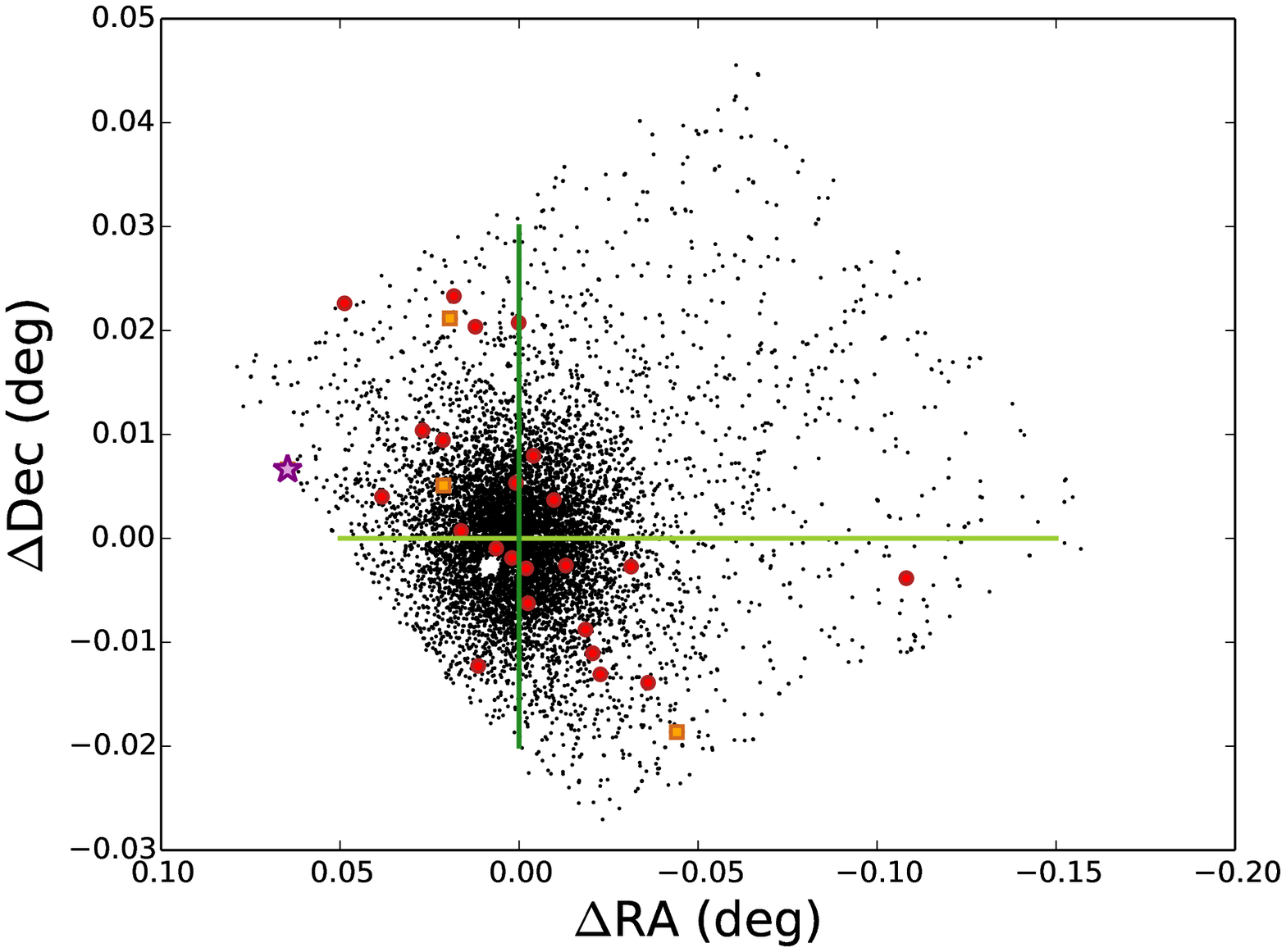}
\caption{Extinction corrected {\it HST}/ACS CMD and stellar map of
  \dwarf. Black and colored symbols indicate the location of detected stars in the ACS field of view
  that are consistent with being in \dwarf. The green line in the left
panel indicates the location of the TRGB as calculated by
\citet{Chiboucas13}. The green lines in the right panel are lines of
PA$=$0 and 90 degrees. We do not find a velocity change
  $>22$~\kms with respect
to these lines or the lines with PA$=$45 or 135 degrees. The three blue lines show stellar isochrones from
Padova models \citep[PARSEC v1.2S$+$COLIBRI PR16]{Bressan12,Marigo13,Rosenfield16} with an age of 12~Gyr and metallicities of
$Z=0.0001$, $Z=0.00055$, and $Z=0.001$ from left to right. The hole
close to the center of the galaxy in the right panel coincides with
the position of a saturated Milky Way star. The red dots indicate
stars that are consistent with being RGB stars. The orange squares
indicate stars consistent with being AGB stars. The purple asterisk
indicates a star consistent with being a blue loop star that is also
blended, due to seeing effects, with a background galaxy.}\label{CMD}
\end{figure*}

We designed a slitmask for the DEIMOS spectrograph \citep{DEIMOS} located at the Keck II 10~m telescope in the Mauna Kea Observatory (Hawaii). We used the color-magnitude diagram (CMD) based on {\it HST}/ACS photometry to select point-like objects that are consistent with being stars in the recently discovered dwarf galaxy \dwarf\ in the halo of M81. We selected the closest stars to the  tip of the red giant branch \citep[$F814W_{TRGB}$=23.77, extinction corrected;][]{Chiboucas13} as spectroscopic targets. Due to the high density of stars, not all the potential targets were observable. Figure \ref{CMD} shows the position of the 27 observed spectroscopic targets in the CMD and their location in \dwarf.

The observations were carried out using the 1200~lines/mm grating centered at 7800~\AA\ with slit widths of 1\farcs0 and the OG550 order blocking filter. All the slits were aligned with the mask position angle (P.A.~$30^{\circ}$). We integrated for a total of 7.2~hours with an average seeing of $0.9''$ on 2016 January $8-9$. This instrumental configuration provides a wavelength coverage of $\sim6500-9000\AA$ with a spectral resolution of $R\sim6000$.

We reduced the data with the {\sc spec2d} pipeline
\citep{DEIMOSpipeline1,DEIMOSpipeline2} with modifications described
by \citet{Kirby15a,Kirby15b}. The major improvements consist of
improving the wavelength solution by tracing the sky lines along the
slit and improving the extraction of the one-dimensional spectra by
accounting for the differential atmospheric refraction along the slit.
The main steps in the reduction process consisted of flat-field corrections, wavelength calibration, sky subtraction, and
cosmic ray cleaning.

\subsection{Observational Biases and Seeing Effects}\label{bias}

We want to use this sample of stars to estimate the radial velocity
and metallicity of \dwarf. To know the reliability of our measurements we study how well we are sampling
the metallicity distribution function of the dwarf and how much light from other sources
contaminates our spectra due to seeing effects.

Our target stars cover the full spatial extent of \dwarf\ as shown in Figure
\ref{CMD}. Thus, if there is a metallicity gradient within the galaxy,
our sample does not favor any particular region. 

We test how well our targets represent the bright end of the RGB
by calculating the perpendicular distance of each star in the
spectroscopic and photometric samples, constrained to the same F814W magnitude range, to the middle isochrone shown in Figure \ref{CMD}. We run
a Kolmogorov-Smirnov statistical test and find that we cannot reject
the null hypothesis of both samples coming from the same parent sample
with a very high confidence ($p-value=0.58$). If we constrain the
parent sample to not only the magnitude but also the color range covered by our targets,
the significance level is even higher ($p-value=0.86$). This means
that we sample the metallicity distribution function of \dwarf\ for $F814W>24.39$.

Our RGBs are selected from {\it HST}/ACS photometry, but
our spectroscopy is ground based and therefore affected by seeing. We
study how much light from neighboring sources 
contribute to our spectra by analyzing the sources that are
within a radius of our typical seeing (${\rm FWHM}=0.9''$) centered on
our targets. We find that $59\%$ of our RGBs have a light contamination
of $0\%$, $37\%$ of our RGBs have a contamination of $<20\%$, and
$4\%$, one star, has a luminosity contamination of $45\%$. This large
contribution comes from a background galaxy whose emission lines are clearly seen
in our spectrum. This galaxy contributes to the continuum of the
spectrum but not to the position of the RGB absorption
lines. Thus, this spectrum can be used to measure a radial velocity
but not a metallicity. In the remaining cases, the light contamination
is null or so low that we do not expect them to affect our
spectroscopic measurements.

\subsection{Background and Foreground Contaminants}\label{cont}

Background galaxies with emission lines are visually
identified and removed from the sample. We detected one that is co-spatial
to an RGB star within our seeing conditions (purple asterisk in Figure
\ref{CMD}). 
Quenched galaxies are not expected within our sample
because of their small radii ($R\sim 1$~kpc for compact quenched
galaxies at redshift $1-2$) with respect to the $\sim0.5$~kpc that {\it HST}/ACS resolves. Smaller
galaxies have typical ${\rm F814W}>26$~mag \citep{vdW14}, too faint to
have been observed.

We use the Besan\c{c}on model \citep{besanconmodel} to estimate the
number of expected Milky Way stars in the line-of-sight of
\dwarf. The model predicts 8~stars with $22<{\rm F814W}<24.5$ and
$0.6<{\rm F606W-F814W}<1.3$ in the {\it HST}/ACS
field-of-view. After applying our specific spectroscopic selection function corresponding to stars near the TRGB, the
expected number of MW stars in our sample is $<1$. Thus, we do not
expect any contaminants within our sample.

We use those regions of the ACS field of view that are furthest from the dwarf to estimate any possible contamination from M81 halo stars, applying our spectroscopic selection function and rescaling the area appropriately, while assuming that any halo contamination at these projected distances from M81 ($\sim$335 kpc) can be approximated by a constant surface density.  At most, we estimate a maximum of one star that could be from M81's halo, and we estimate the effects of this potential small contamination in Section~\ref{vel}.

\section{Spectroscopic Measurements}

\subsection{Radial Velocity}\label{vel}

\begin{figure}
\centering
\includegraphics[angle=0,width=9cm]{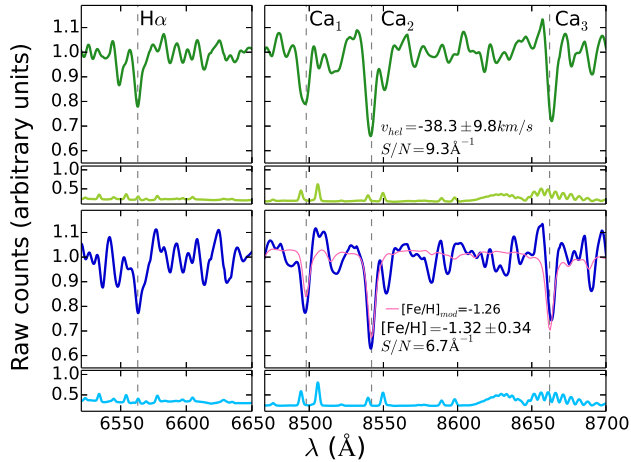}
\caption{Spectra showing the H$\alpha$ and calcium triplet
  lines. Upper panels: flux (dark green) and error (light green) spectra that result after coadding all 27
  targets used to measure the radial velocity of
  \dwarf. Lower panels: flux (dark blue) and error (light blue) spectra that result after
  coadding the 23 targets used to estimate the metallicity (red dots
  in Figure \ref{CMD}). The pink line is a MIUSCAT model of a single stellar population
  with age of 14~Gyr and the indicated metallicity for comparison
  purposes \citep{Vazdekis12}. The weak Ca$_1$ line is not used to
  estimate the metallicity \citep[see][]{Ho15}. }\label{spectra}
\end{figure}

Due to the faintness of the targeted stars, they do not have
enough identifiable absorption lines in the individual spectra to obtain a reliable
line-of-sight radial velocity ($v$). To improve the reliability of $v$
we coadd all 27 stars together (see Figure \ref{spectra}). Our spectral coaddition process is
the same one as described by \citet{etj16c}: (1) correct for
possible offsets across the slit using the atmospheric A band seen in
the continuum of all objects, this correction is $<10$~\kms; (2) rebin
the spectra and their uncertainties to a common wavelength range; (3)
renormalize the fluxes and their associated uncertainties; and (4)
add, pixel by pixel, the fluxes of the renormalized rebinned spectra
by performing a sigma clipping where those pixels that deviate more than $3\sigma$
from the median are rejected. 

The $v$ of this coadded spectrum is measured using the penalized
pixel-fitting (pPXF) method developed by \citet[][]{PPXF}. This software
finds the composite stellar template that best fits our coadded
spectrum. The composite stellar template is a linear combination of
the stars in our stellar library (see below) allowing for different weights to
minimize template mismatch. 
Our stellar library consists of 9 high signal-to-noise (S/N~$>100$) stars of different spectral types
(A$-$K), luminosity classes (I$-$V), and metallicities ($-3<{\rm
  [Fe/H]<0}$) that were observed with the same instrumental
configuration.

We computed the uncertainty due to random noise by finding the
standard deviation of the velocities of 1000 Monte Carlo realizations
of the spectrum. In each realization the spectrum is perturbed pixel
by pixel within a Gaussian
function whose width is the flux uncertainty. To this random
uncertainty we add in quadrature a systematic uncertainty of
$1.49$~\kms\ as estimated by \citet{Kirby15b}. This systematic
uncertainty, estimated by comparing repeated measurements of the same
stars, includes effects such as uncorrected spectrograph flexure
or small errors in the wavelength solution. The measured heliocentric
velocity of \dwarf~is $-38.3\pm9.8$~\kms, as shown in Table \ref{data}.

In Section \ref{cont} we determined that we expect to have a maximum
  of one star from M81's halo contaminating our sample. We estimate
  the effect that such a star would
have on our velocity measurement. The line-of-sight heliocentric
velocity of M81 is $-36$~\kms\ and the velocity dispersion in the
outer halo is 84~\kms\ \citep{Karachentsev02}. We make two tests: (1)
we calculate the line-of-sight velocity after dropping one
star at a time from our coadded spectrum and (2) we carry out 100
simulations where we add a random
velocity offset within a Gaussian function whose dispersion is 84~\kms\ to
one star in the coadded spectrum. The dispersion of the 27
velocity measurements for test (1) is 1.7~\kms\ and of the 100
simulations of test (2) is 8.7~\kms. Such small dispersions suggest
that a single potential M81 contaminant has
no effect on our velocity measurement.

We search for velocity gradients in four different directions (using
trial position angles of PA$=$0, 45, 90, and 135~degrees) given
the roundness of this object (see its ellipticity in Table
\ref{data}). After splitting the velocities into two groups for each
trial PA, and searching for differences in velocity between them, no
significant gradients were found within the uncertainties. Our
  data rule out a maximum velocity change of $>22$~\kms\ and a
velocity gradient of $>345$~km~s$^{-1}$~kpc$^{-1}$. However, the expected values
based on Local Group dSphs of similar luminosity are below our limits \citep{Aden09,Ho12,Collins16}, and so this is not a strong constraint.

\subsection{Metallicity}

We estimate the stellar metallicity of \dwarf\ following the same
procedure as \citet{Ho15} and \citet{etj16c}. This method transforms the
equivalent width (EW) of the two strongest absorption lines in the
calcium triplet into [Fe/H]. These two lines are fitted with a
Gaussian function. The measured EW is
transformed into the EW value that would have been obtained if we had fitted a Gaussian plus a Lorentzian function. The obtained
values are then converted into the total EW of the calcium lines by
making an unweighted sum $\Sigma {\rm Ca} ={\rm EW_{8542}}+{\rm
  EW_{8662}}$, and then transformed into [Fe/H] by using the
calibration of \citet{Carrera13}, also used by \citet{Ho15}:

\begin{eqnarray}\label{metallicity}
{\rm [Fe/H]} &=& -3.51+0.12\times M_I+0.57\times \Sigma{\rm Ca} \nonumber \\
                   & -& 0.17\times \Sigma{\rm Ca}^{-1.5} +0.02\times \Sigma{\rm Ca}\times M_I
\end{eqnarray}

\noindent where $M_I$ is the absolute $I$-band magnitude. The uncertainty in [Fe/H] is 
the propagation of the uncertainties in ${\rm \Sigma Ca}$ and $M_I$
accounting for the photometric errors of the $I$-band magnitudes of
the individual stars and the measured distance to \dwarf\ by
\citet{Chiboucas13}. 

Due to the low S/N of the individual spectra, we coadd them
together. \citet{Yang13} demonstrated that when
coadding the stars the metallicity measured is the average
[Fe/H] of the group of stars. As the calibration above is calculated
for RGB stars, we coadd those 23 targets that are consistent with being RGBs (red dots in
Figure \ref{CMD}, see the resulting spectrum in Figure \ref{spectra}). $M_I$, in this case, is the average absolute $I$-band magnitude of
the 23 stars in the coadded spectrum. 

We find a  metallicity of [Fe/H] = $-$1.3$\pm$0.3, listed in Table
\ref{data}. It is not possible to measure a metallicity gradient due
to the low S/N of the spectra. The photometric metallicity estimated using the
full population of detected stars agrees well with the spectroscopic
value ([Fe/H]$_{phot}=-1.3\pm0.4$). The estimated photometric
metallicity gradient is very mild $d{\rm
  [Fe/H]}/(r/re)\sim0.06$~dex$/re$.

Following the same calculations as in Section \ref{vel} we
  estimate the effect on our measured metallicity if
  one M81 halo star were contaminating our sample. The
  dispersion of the 23 values of [Fe/H] measured after dropping one star at a
  time from our coadded spectrum is 0.2~dex. However, shifting the
  velocity of one of the 23 coadded spectra within the dispersion of
  M81's outer halo introduces an offset in the metallicity towards
  more metal poor values. We account for these effects in the
  uncertainties of our quoted metallicity.

\begin{table}
\begin{center}
\caption{Properties of \dwarf \label{data}}
{\renewcommand{\arraystretch}{1.}
\begin{tabular}{l|c}
\hline \hline
Parameter         &  Value  \\
\hline
RA$_0$~(hh:mm:ss)$^{(1)}$   & 9:44:34.37       \\
DEC$_0$~(dd:mm:ss)$^{(1)}$ & 71:28:55.60    \\
$m-M$~(mag)$^{(1)}$         & $27.87\pm_{0.22}^{0.32}$ \\
$D$~(Mpc)$^{(1)}$               & $3.7\pm_{0.4}^{0.5}$  \\
$M_r$~(mag)$^{(1)}$           & $-12.4\pm0.8$ \\
$M_I$~(mag)$^{(1)}$           & $-13.2\pm0.4$       \\
$M_{HI}$ ($M_{\odot}$)$^{(2)}$  & $<$3.1$\times$10$^5$\\
$(V-I)_{TRGB}$$^{(1)}$           & $1.13\pm0.07$ \\
$r_e$~(arcsec)$^{(1)}$         & $21.4\pm 0.4$ \\
$r_e$~(kpc)$^{(1)}$                & $0.35\pm0.02$ \\
$\epsilon$$^{(1)}$               & $0.11$ \\
PA~(N to E; deg)$^{(1)}$       & $3.6$ \\
$\mu_{0,r}$~(mag~arcsec$^{-2}$)$^{(1)}$ &  $23.4$   \\
$\langle\mu_{e,r}\rangle$~(mag~arcsec$^{-2}$)$^{(1)}$ &   $23.9$  \\
$D$(M81-\dwarf)~(deg)$^{(3)}$ & 2.59 \\
$D$(M81-\dwarf)~(Mpc)$^{(3)}$ & 0.335 \\
\hline
$v_{hel}$~\kms$^{(4)}$ &$-38.3\pm9.8$\\
$\Sigma Ca$$^{(4)}$    & $5.4\pm0.6$\\
${\rm [Fe/H]}$~(dex)$^{(4)}$     & $-1.3^{+0.3}_{-0.6}$\\
\hline
\end{tabular}}
\end{center}
\tablecomments{Rows 1-16 are parameters taken from the
  literature. Rows 17-19 are derived in this paper.\\
$^{(1)}$ Parameters from \citet{Chiboucas13}. The
 central coordinates of the galaxy are in J2000. The magnitudes are
  measured in the AB system. $M_r$ is measured from ground-based
  CFHT/MegaCam photometry and $M_I$ from {\it HST} photometry. The color $V-I$ is calculated in the
  TRGB. The parameter $r_e$ is the half-light radius, and $\mu_{0,r}$  and
  $\langle\mu_{h,r}\rangle$ are the central and effective surface
  brightness. \\
$^{(2)}$ From \citet{Roychowdhury12}.\\
$^{(3)}$ Projected distance from M81 to \dwarf. \\
$^{(4)}$ Heliocentric
velocity, total EW of the calcium triplet, and metallicity measured in
this work.
}
\end{table}

\section{Discussion and Conclusions}\label{discussion}

\begin{figure}
\centering
\includegraphics[angle=0,width=9cm]{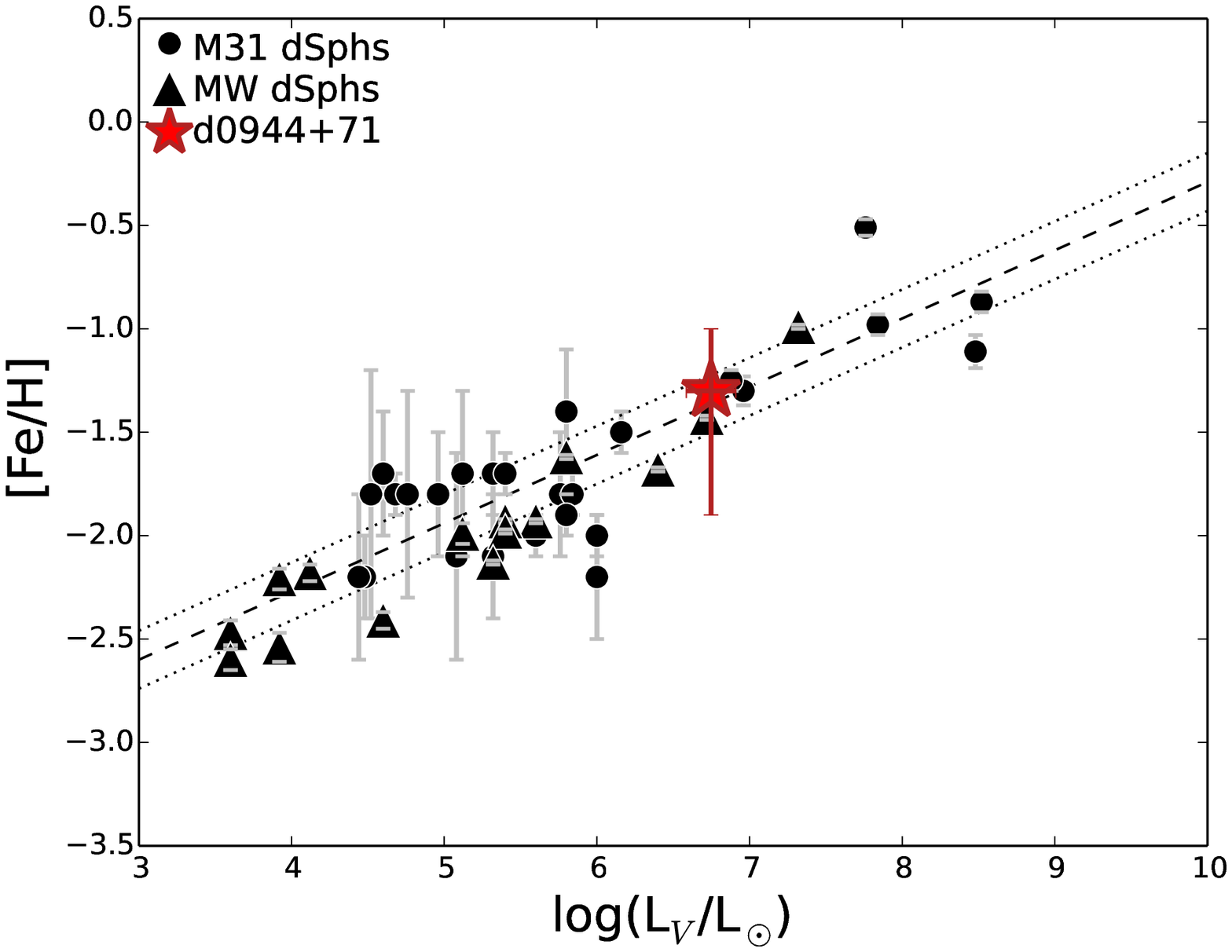}
\caption{Metallicity$-$luminosity relation for LG dSphs and
  \dwarf. The measurements for the MW dSphs were done by \citet{Kirby11} and
  for M31 dSphs by \citet{Collins13,Collins15,Ho15}.}\label{metallicity-luminosity}
\end{figure}

We use Keck/DEIMOS spectroscopy to measure the line-of-sight radial
velocity and the stellar metallicity of \dwarf, a satellite of
the spiral galaxy M81. Due to the faintness of the targeted RGB stars 
($22.63<{\rm F814W}<24.39$) we cannot make spectroscopic measurements
in individual stars, but, we can coadd them to obtain average
measurements for the dSph. 

We coadd a total of 27 stars in \dwarf\ and measure a heliocentric
radial velocity of $-$38.3$\pm$9.8~\kms.
Including or removing those stars not consistent with
being RGB stars in our sample does not change our measured velocities
within the uncertainties. This heliocentric velocity is consistent
with the velocity of M81 \citep[$-36$~\kms;][]{Karachentsev02}, which
is 0.34~Mpc away in projection. This
suggests that \dwarf\ is likely gravitationally bound to
M81.

We coadd 23 RGB stars that span $\sim0.8$~mag below the TRGB to
estimate a metallicity of [Fe/H]$=-1.3\pm0.3$ based on the calcium
triplet lines. The low S/N of our
spectra does not allow us to measure a metallicity gradient. This
metallicity is consistent with that of Leo~I, And~VII, and And~XVIII,
the three Local Group dSphs that have similar luminosity to \dwarf\
\citep{Kirby11,Collins13,Ho15}. 

Figure
\ref{metallicity-luminosity} shows that \dwarf\ is consistent with the metallicity$-$luminosity
relation for dSphs in the Local Group.
Outliers above the metallicity$-$luminosity relation are likely
galaxies that have been tidally disturbed. In such an event, the
galaxy loses stars becoming fainter but keeps the same metallicity
if new star formation is not triggered.
Outliers below the metallicity$-$luminosity relation could be
explained by internal metallicity gradients. A galaxy with a more
metal-poor population in the  outskirts would have a
more metal-poor measurement than what would have been obtained from a
sample of centrally concentrated RGBs. The position of this dSph in
the metallicity$-$luminosity relation in combination with its lack of
metallicity gradient measured by the colors of its stars, its
roundness, and its lack of tidal
disturbances in the spatial distribution of the stars suggest that \dwarf\ has not suffered major tidal
stripping.

This is the first of a series of papers where we will apply this powerful
technique to study the dynamical
and metallicity properties of dSph galaxies beyond the Local Group. We
will target other galaxies in the M81 group to analyze their orbital
properties, the mass of M81, the possible plane where all the
satellites lie, and search for dynamical differences between the
population of dSphs and dwarf irregulars.

\acknowledgments

E.T. and P.G. acknowledge the NSF grants AST-1010039 and
AST-1412504. D.J.S. is supported by NSF grants AST-1412504 and
AST-1517649 and J.D.S. by AST-1412792.  The work of DJS was partially performed at the Aspen Center for Physics, which is supported by National Science Foundation grant PHY-1066293.
The spectroscopic data presented herein were obtained at the
W.M. Keck Observatory, which is operated as a scientific partnership
among the California Institute of Technology, the University of
California and the National Aeronautics and Space
Administration. The Observatory was made possible by the generous
financial support of the W.M. Keck Foundation. The authors
wish to recognize and acknowledge the very significant cultural role
and reverence that the summit of Mauna Kea has always had within the
indigenous Hawaiian community. We are most fortunate to have the
opportunity to conduct observations from this mountain.

\bibliographystyle{aa}
\bibliography{references}{}

\begin{thebibliography}{47}
\expandafter\ifx\csname natexlab\endcsname\relax\def\natexlab#1{#1}\fi

\bibitem[{{Ad{\'e}n} {et~al.}(2009){Ad{\'e}n}, {Wilkinson}, {Read}, {Feltzing},
  {Koch}, {Gilmore}, {Grebel}, \& {Lundstr{\"o}m}}]{Aden09}
{Ad{\'e}n}, D., {Wilkinson}, M.~I., {Read}, J.~I., {et~al.} 2009, \apjl, 706,
  L150

\bibitem[{{Boylan-Kolchin} {et~al.}(2011){Boylan-Kolchin}, {Bullock}, \&
  {Kaplinghat}}]{BoylanKolchin11}
{Boylan-Kolchin}, M., {Bullock}, J.~S., \& {Kaplinghat}, M. 2011, \mnras, 415,
  L40

\bibitem[{{Boylan-Kolchin} {et~al.}(2012){Boylan-Kolchin}, {Bullock}, \&
  {Kaplinghat}}]{BoylanKolchin12}
{Boylan-Kolchin}, M., {Bullock}, J.~S., \& {Kaplinghat}, M. 2012, \mnras, 422,
  1203

\bibitem[{{Bressan} {et~al.}(2012){Bressan}, {Marigo}, {Girardi}, {Salasnich},
  {Dal Cero}, {Rubele}, \& {Nanni}}]{Bressan12}
{Bressan}, A., {Marigo}, P., {Girardi}, L., {et~al.} 2012, \mnras, 427, 127

\bibitem[{{Brooks} \& {Zolotov}(2014)}]{Brooks14}
{Brooks}, A.~M. \& {Zolotov}, A. 2014, \apj, 786, 87

\bibitem[{{Busha} {et~al.}(2010){Busha}, {Alvarez}, {Wechsler}, {Abel}, \&
  {Strigari}}]{Busha10}
{Busha}, M.~T., {Alvarez}, M.~A., {Wechsler}, R.~H., {Abel}, T., \& {Strigari},
  L.~E. 2010, \apj, 710, 408

\bibitem[{{Cappellari} \& {Emsellem}(2004)}]{PPXF}
{Cappellari}, M. \& {Emsellem}, E. 2004, \pasp, 116, 138

\bibitem[{{Carlin} {et~al.}(2016){Carlin}, {Sand}, {Price}, {Willman},
  {Karunakaran}, {Spekkens}, {Bell}, {Brodie}, {Crnojevi{\'c}}, {Forbes},
  {Hargis}, {Kirby}, {Lupton}, {Peter}, {Romanowsky}, \& {Strader}}]{Carlin16}
{Carlin}, J.~L., {Sand}, D.~J., {Price}, P., {et~al.} 2016, \apjl, 828, L5

\bibitem[{{Carrera} {et~al.}(2013){Carrera}, {Pancino}, {Gallart}, \& {del
  Pino}}]{Carrera13}
{Carrera}, R., {Pancino}, E., {Gallart}, C., \& {del Pino}, A. 2013, \mnras,
  434, 1681

\bibitem[{{Chiboucas} {et~al.}(2013){Chiboucas}, {Jacobs}, {Tully}, \&
  {Karachentsev}}]{Chiboucas13}
{Chiboucas}, K., {Jacobs}, B.~A., {Tully}, R.~B., \& {Karachentsev}, I.~D.
  2013, \aj, 146, 126

\bibitem[{{Chiboucas} {et~al.}(2009){Chiboucas}, {Karachentsev}, \&
  {Tully}}]{Chiboucas09}
{Chiboucas}, K., {Karachentsev}, I.~D., \& {Tully}, R.~B. 2009, \aj, 137, 3009

\bibitem[{{Collins} {et~al.}(2013){Collins}, {Chapman}, {Rich}, {Ibata},
  {Martin}, {Irwin}, {Bate}, {Lewis}, {Pe{\~n}arrubia}, {Arimoto}, {Casey},
  {Ferguson}, {Koch}, {McConnachie}, \& {Tanvir}}]{Collins13}
{Collins}, M.~L.~M., {Chapman}, S.~C., {Rich}, R.~M., {et~al.} 2013, \apj, 768,
  172

\bibitem[{{Collins} {et~al.}(2015){Collins}, {Martin}, {Rich}, {Ibata},
  {Chapman}, {McConnachie}, {Ferguson}, {Irwin}, \& {Lewis}}]{Collins15}
{Collins}, M.~L.~M., {Martin}, N.~F., {Rich}, R.~M., {et~al.} 2015, \apjl, 799,
  L13

\bibitem[{{Collins} {et~al.}(2016){Collins}, {Tollerud}, {Sand}, {Bonaca},
  {Willman}, \& {Strader}}]{Collins16}
{Collins}, M.~L.~M., {Tollerud}, E.~J., {Sand}, D.~J., {et~al.} 2016, ArXiv
  e-prints

\bibitem[{{Cooper} {et~al.}(2012){Cooper}, {Newman}, {Davis}, {Finkbeiner}, \&
  {Gerke}}]{DEIMOSpipeline1}
{Cooper}, M.~C., {Newman}, J.~A., {Davis}, M., {Finkbeiner}, D.~P., \& {Gerke},
  B.~F. 2012, {spec2d: DEEP2 DEIMOS Spectral Pipeline}, astrophysics Source
  Code Library, ascl:1203.003

\bibitem[{{Crnojevi{\'c}} {et~al.}(2014){Crnojevi{\'c}}, {Sand}, {Caldwell},
  {Guhathakurta}, {McLeod}, {Seth}, {Simon}, {Strader}, \&
  {Toloba}}]{Crnojevic14}
{Crnojevi{\'c}}, D., {Sand}, D.~J., {Caldwell}, N., {et~al.} 2014, \apjl, 795,
  L35

\bibitem[{{Crnojevi{\'c}} {et~al.}(2016){Crnojevi{\'c}}, {Sand}, {Spekkens},
  {Caldwell}, {Guhathakurta}, {McLeod}, {Seth}, {Simon}, {Strader}, \&
  {Toloba}}]{Crnojevic16}
{Crnojevi{\'c}}, D., {Sand}, D.~J., {Spekkens}, K., {et~al.} 2016, \apj, 823,
  19

\bibitem[{{Faber} {et~al.}(2003){Faber}, {Phillips}, {Kibrick}, {Alcott},
  {Allen}, {Burrous}, {Cantrall}, {Clarke}, {Coil}, {Cowley}, {Davis}, {Deich},
  {Dietsch}, {Gilmore}, {Harper}, {Hilyard}, {Lewis}, {McVeigh}, {Newman},
  {Osborne}, {Schiavon}, {Stover}, {Tucker}, {Wallace}, {Wei}, {Wirth}, \&
  {Wright}}]{DEIMOS}
{Faber}, S.~M., {Phillips}, A.~C., {Kibrick}, R.~I., {et~al.} 2003, in Society
  of Photo-Optical Instrumentation Engineers (SPIE) Conference Series, Vol.
  4841, Instrument Design and Performance for Optical/Infrared Ground-based
  Telescopes, ed. M.~{Iye} \& A.~F.~M. {Moorwood}, 1657--1669

\bibitem[{{Ho} {et~al.}(2012){Ho}, {Geha}, {Munoz}, {Guhathakurta}, {Kalirai},
  {Gilbert}, {Tollerud}, {Bullock}, {Beaton}, \& {Majewski}}]{Ho12}
{Ho}, N., {Geha}, M., {Munoz}, R.~R., {et~al.} 2012, \apj, 758, 124

\bibitem[{{Ho} {et~al.}(2015){Ho}, {Geha}, {Tollerud}, {Zinn}, {Guhathakurta},
  \& {Vargas}}]{Ho15}
{Ho}, N., {Geha}, M., {Tollerud}, E.~J., {et~al.} 2015, \apj, 798, 77

\bibitem[{{Ibata} {et~al.}(2013){Ibata}, {Lewis}, {Conn}, {Irwin},
  {McConnachie}, {Chapman}, {Collins}, {Fardal}, {Ferguson}, {Ibata}, {Mackey},
  {Martin}, {Navarro}, {Rich}, {Valls-Gabaud}, \& {Widrow}}]{Ibata13}
{Ibata}, R.~A., {Lewis}, G.~F., {Conn}, A.~R., {et~al.} 2013, \nat, 493, 62

\bibitem[{{Johnston} {et~al.}(2008){Johnston}, {Bullock}, {Sharma}, {Font},
  {Robertson}, \& {Leitner}}]{Johnston08}
{Johnston}, K.~V., {Bullock}, J.~S., {Sharma}, S., {et~al.} 2008, \apj, 689,
  936

\bibitem[{{Karachentsev} {et~al.}(2002){Karachentsev}, {Dolphin}, {Geisler},
  {Grebel}, {Guhathakurta}, {Hodge}, {Karachentseva}, {Sarajedini}, {Seitzer},
  \& {Sharina}}]{Karachentsev02}
{Karachentsev}, I.~D., {Dolphin}, A.~E., {Geisler}, D., {et~al.} 2002, \aap,
  383, 125

\bibitem[{{Kirby} {et~al.}(2015{\natexlab{a}}){Kirby}, {Guo}, {Zhang}, {Deng},
  {Cohen}, {Guhathakurta}, {Shetrone}, {Lee}, \& {Rizzi}}]{Kirby15a}
{Kirby}, E.~N., {Guo}, M., {Zhang}, A.~J., {et~al.} 2015{\natexlab{a}}, \apj,
  801, 125

\bibitem[{{Kirby} {et~al.}(2011){Kirby}, {Lanfranchi}, {Simon}, {Cohen}, \&
  {Guhathakurta}}]{Kirby11}
{Kirby}, E.~N., {Lanfranchi}, G.~A., {Simon}, J.~D., {Cohen}, J.~G., \&
  {Guhathakurta}, P. 2011, \apj, 727, 78

\bibitem[{{Kirby} {et~al.}(2015{\natexlab{b}}){Kirby}, {Simon}, \&
  {Cohen}}]{Kirby15b}
{Kirby}, E.~N., {Simon}, J.~D., \& {Cohen}, J.~G. 2015{\natexlab{b}}, \apj,
  810, 56

\bibitem[{{Klypin} {et~al.}(1999){Klypin}, {Kravtsov}, {Valenzuela}, \&
  {Prada}}]{Klypin99}
{Klypin}, A., {Kravtsov}, A.~V., {Valenzuela}, O., \& {Prada}, F. 1999, \apj,
  522, 82

\bibitem[{{Marigo} {et~al.}(2013){Marigo}, {Bressan}, {Nanni}, {Girardi}, \&
  {Pumo}}]{Marigo13}
{Marigo}, P., {Bressan}, A., {Nanni}, A., {Girardi}, L., \& {Pumo}, M.~L. 2013,
  \mnras, 434, 488

\bibitem[{{Mart{\'{\i}}nez-Delgado} {et~al.}(2015){Mart{\'{\i}}nez-Delgado},
  {D'Onghia}, {Chonis}, {Beaton}, {Teuwen}, {GaBany}, {Grebel}, \&
  {Morales}}]{MD15}
{Mart{\'{\i}}nez-Delgado}, D., {D'Onghia}, E., {Chonis}, T.~S., {et~al.} 2015,
  \aj, 150, 116

\bibitem[{{Merritt} {et~al.}(2014){Merritt}, {van Dokkum}, \&
  {Abraham}}]{Merritt14}
{Merritt}, A., {van Dokkum}, P., \& {Abraham}, R. 2014, \apjl, 787, L37

\bibitem[{{Moore} {et~al.}(1999){Moore}, {Ghigna}, {Governato}, {Lake},
  {Quinn}, {Stadel}, \& {Tozzi}}]{Moore99}
{Moore}, B., {Ghigna}, S., {Governato}, F., {et~al.} 1999, \apjl, 524, L19

\bibitem[{{Newman} {et~al.}(2013){Newman}, {Cooper}, {Davis}, {Faber}, {Coil},
  {Guhathakurta}, {Koo}, {Phillips}, {Conroy}, {Dutton}, {Finkbeiner}, {Gerke},
  {Rosario}, {Weiner}, {Willmer}, {Yan}, {Harker}, {Kassin}, {Konidaris},
  {Lai}, {Madgwick}, {Noeske}, {Wirth}, {Connolly}, {Kaiser}, {Kirby},
  {Lemaux}, {Lin}, {Lotz}, {Luppino}, {Marinoni}, {Matthews}, {Metevier}, \&
  {Schiavon}}]{DEIMOSpipeline2}
{Newman}, J.~A., {Cooper}, M.~C., {Davis}, M., {et~al.} 2013, \apjs, 208, 5

\bibitem[{{Pawlowski} {et~al.}(2015){Pawlowski}, {McGaugh}, \&
  {Jerjen}}]{Pawlowski15}
{Pawlowski}, M.~S., {McGaugh}, S.~S., \& {Jerjen}, H. 2015, \mnras, 453, 1047

\bibitem[{{Robin} {et~al.}(2003){Robin}, {Reyl{\'e}}, {Derri{\`e}re}, \&
  {Picaud}}]{besanconmodel}
{Robin}, A.~C., {Reyl{\'e}}, C., {Derri{\`e}re}, S., \& {Picaud}, S. 2003,
  \aap, 409, 523

\bibitem[{{Rosenfield} {et~al.}(2016){Rosenfield}, {Marigo}, {Girardi},
  {Dalcanton}, {Bressan}, {Williams}, \& {Dolphin}}]{Rosenfield16}
{Rosenfield}, P., {Marigo}, P., {Girardi}, L., {et~al.} 2016, \apj, 822, 73

\bibitem[{{Roychowdhury} {et~al.}(2012){Roychowdhury}, {Chengalur},
  {Chiboucas}, {Karachentsev}, {Tully}, \& {Kaisin}}]{Roychowdhury12}
{Roychowdhury}, S., {Chengalur}, J.~N., {Chiboucas}, K., {et~al.} 2012, \mnras,
  426, 665

\bibitem[{{Sand} {et~al.}(2014){Sand}, {Crnojevi{\'c}}, {Strader}, {Toloba},
  {Simon}, {Caldwell}, {Guhathakurta}, {McLeod}, \& {Seth}}]{Sand14}
{Sand}, D.~J., {Crnojevi{\'c}}, D., {Strader}, J., {et~al.} 2014, \apjl, 793,
  L7

\bibitem[{{Sand} {et~al.}(2015){Sand}, {Spekkens}, {Crnojevi{\'c}}, {Hargis},
  {Willman}, {Strader}, \& {Grillmair}}]{Sand15}
{Sand}, D.~J., {Spekkens}, K., {Crnojevi{\'c}}, D., {et~al.} 2015, \apjl, 812,
  L13

\bibitem[{{Toloba} {et~al.}(2016{\natexlab{a}}){Toloba}, {Guhathakurta},
  {Romanowsky}, {Brodie}, {Martinez-Delgado}, {Arnold}, {Ramachandran}, \&
  {Theakanath}}]{etj16c}
{Toloba}, E., {Guhathakurta}, P., {Romanowsky}, A., {et~al.}
  2016{\natexlab{a}}, ArXiv e-prints

\bibitem[{{Toloba} {et~al.}(2016{\natexlab{b}}){Toloba}, {Sand}, {Spekkens},
  {Crnojevi{\'c}}, {Simon}, {Guhathakurta}, {Strader}, {Caldwell}, {McLeod}, \&
  {Seth}}]{etj16a}
{Toloba}, E., {Sand}, D.~J., {Spekkens}, K., {et~al.} 2016{\natexlab{b}},
  \apjl, 816, L5

\bibitem[{{van der Wel} {et~al.}(2014){van der Wel}, {Franx}, {van Dokkum},
  {Skelton}, {Momcheva}, {Whitaker}, {Brammer}, {Bell}, {Rix}, {Wuyts},
  {Ferguson}, {Holden}, {Barro}, {Koekemoer}, {Chang}, {McGrath},
  {H{\"a}ussler}, {Dekel}, {Behroozi}, {Fumagalli}, {Leja}, {Lundgren},
  {Maseda}, {Nelson}, {Wake}, {Patel}, {Labb{\'e}}, {Faber}, {Grogin}, \&
  {Kocevski}}]{vdW14}
{van der Wel}, A., {Franx}, M., {van Dokkum}, P.~G., {et~al.} 2014, \apj, 788,
  28

\bibitem[{{van Dokkum} {et~al.}(2016){van Dokkum}, {Abraham}, {Brodie},
  {Conroy}, {Danieli}, {Merritt}, {Mowla}, {Romanowsky}, \&
  {Zhang}}]{vandokkum16}
{van Dokkum}, P., {Abraham}, R., {Brodie}, J., {et~al.} 2016, ArXiv e-prints

\bibitem[{{van Dokkum} {et~al.}(2015){van Dokkum}, {Romanowsky}, {Abraham},
  {Brodie}, {Conroy}, {Geha}, {Merritt}, {Villaume}, \& {Zhang}}]{vandokkum15}
{van Dokkum}, P.~G., {Romanowsky}, A.~J., {Abraham}, R., {et~al.} 2015, \apjl,
  804, L26

\bibitem[{{Vazdekis} {et~al.}(2012){Vazdekis}, {Ricciardelli}, {Cenarro},
  {Rivero-Gonz{\'a}lez}, {D{\'{\i}}az-Garc{\'{\i}}a}, \&
  {Falc{\'o}n-Barroso}}]{Vazdekis12}
{Vazdekis}, A., {Ricciardelli}, E., {Cenarro}, A.~J., {et~al.} 2012, \mnras,
  424, 157

\bibitem[{{Weisz} {et~al.}(2014){Weisz}, {Dolphin}, {Skillman}, {Holtzman},
  {Gilbert}, {Dalcanton}, \& {Williams}}]{Weisz14}
{Weisz}, D.~R., {Dolphin}, A.~E., {Skillman}, E.~D., {et~al.} 2014, \apj, 789,
  148

\bibitem[{{Wetzel} {et~al.}(2016){Wetzel}, {Hopkins}, {Kim}, {Faucher-Giguere},
  {Keres}, \& {Quataert}}]{Wetzel16}
{Wetzel}, A.~R., {Hopkins}, P.~F., {Kim}, J.-h., {et~al.} 2016, ArXiv e-prints

\bibitem[{{Yang} {et~al.}(2013){Yang}, {Kirby}, {Guhathakurta}, {Peng}, \&
  {Cheng}}]{Yang13}
{Yang}, L., {Kirby}, E.~N., {Guhathakurta}, P., {Peng}, E.~W., \& {Cheng}, L.
  2013, \apj, 768, 4

\end{thebibliography}


\end{document}